\documentstyle[aps,pre,12pt]{revtex}          
\input epsf
\tighten                                  
\begin{document}                          
\draft                                    

\makeatletter
\def\gsim{\compoundrel>\over\sim}
\def\lsim{\compoundrel<\over\sim}
\def\compoundrel#1\over#2{\mathpalette\compoundreL{{#1}\over{#2}}}
\def\compoundreL#1#2{\compoundREL#1#2}
\def\compoundREL#1#2\over#3{\mathrel
      {\vcenter{\hbox{$\m@th\buildrel{#1#2}\over{#1#3}$}}}}
\makeatother

\title{Dynamics of Chainlike Molecules on Surfaces}

\author{T. Hjelt$^{1}$, S. Herminghaus$^{2}$, 
T. Ala-Nissila$^{1,3}$, and S.C. Ying$^{3}$}

\address{
$^1$Helsinki Institute of Physics,
P.O. Box 9, FIN--00014 University of Helsinki, 
Helsinki, Finland \\
}

\address{
$^2$Max-Planck-Institut f\"ur Kolloid- und Grenzfl\"achenforschung, Rudower
Chaussee 5, D--12489 Berlin, Germany\\
}

\address{
$^3$Department of Physics, Box 1843, Brown University,
Providence, R.I. 02912 \\
}

\date{August 15, 1997}

\maketitle

\begin{abstract}
We consider the diffusion and spreading of chainlike molecules on
solid surfaces. We first show that the steep spherical cap shape density
profiles, observed in some submonolayer 
experiments on spreading polymer films, imply
that the collective diffusion coefficient $D_C(\theta)$ must be an
increasing function of the surface coverage $\theta$ for small and 
intermediate coverages.
Through simulations of a discrete model of interacting chainlike molecules,
we demonstrate that this is caused by an 
entropy-induced repulsive interaction. Excellent agreement is found
between experimental and numerically obtained density profiles in
this case, demonstrating that steep submonolayer film edges
naturally arise due to the diffusive properties of
chainlike molecules.
When the entropic repulsion dominates over
interchain attractions, $D_C(\theta)$ first increases as a function
of $\theta$ but then eventually approaches zero for $\theta \rightarrow 1$.
The maximum value of $D_C(\theta)$ decreases for increasing
attractive interactions, leading to density profiles
that are in between spherical cap and Gaussian shapes. 
We also develop an analytic mean field
approach to explain the diffusive behavior
of chainlike molecules.
The thermodynamic factor in $D_C(\theta)$ is
evaluated using effective free energy arguments, and
the chain mobility is calculated numerically using the recently
developed dynamic mean field theory. Good agreement
is obtained between theory and simulations. 
\end{abstract}


\section{Introduction}

The dynamics of polymers on solid surfaces is an
interesting theoretical problem that has
important applications related to thin surface films.
A central role in such systems is
played by the diffusive dynamics of the chains. While  
the diffusion of adatoms and small molecules
on surfaces is an extensively studied problem \cite{Gom90,Ala92},
there are relatively few studies of the diffusion of more
complicated molecules, and especially polymers 
on surfaces \cite{George,Coh92,Fic94}. 
The pioneering experiments of George
{\it et al.} \cite{George} on short-chain alcanes on metal surfaces
revealed that the coverage dependence of the collective
diffusion coefficient $D_C(\theta)$ may show unusual
features. For some molecules, it is approximately constant
while for others, it is an increasing function of coverage for
a wide range of values of $\theta$. 

There has also been substantial interest recently
on the spreading dynamics of molecularly  
thin oil films on solid substrates \cite{Hes89,Caz90,Alb92,Fra93,Haa95}.
Many of these experiments are performed by depositing tiny, very
flat droplets on surfaces. 
In the limit where the film becomes less than one monolayer in
thickness, the spreading molecules are all in 
contact with surface forming a 2D molecular gas.
In this regime, the dynamics of spreading and consequently the
density profiles of the film edges are determined by the diffusion
of the molecules. Experimental studies of
spreading in the submonolayer regime
\cite{Caz90,Alb92,Fra93} reveal that most of the  
measured film edge profiles are not well approximated by the 
Gaussian function but assume
a steeper shape \cite{Her95} that can be well fitted by a spherical cap
in the rotationally symmetric case, {\it i.e.} droplet spreading
\cite{Alb92,Fra93}. 

In the regime where the diffusion equation (Fick's law) is
a valid description, a non-Gaussian profile
indicates that the collective diffusion 
constant $D_C(\theta)$ must have nontrivial coverage dependence.
In the present context this was first realized in conjunction with
the experiments of Albrecht {\it et al.}
\cite{Alb92}, where the observed
spherical cap shaped droplets could be reproduced by Monte Carlo
simulations of spreading of flexible chains in 2D. Analyzed in terms of
Fick's law with a coverage-dependent $D_c(\theta)$, it was
concluded in Ref. \cite{Her95} that $D_c(\theta)$ is a strongly
{\it increasing} function of coverage up to $\theta \approx 0.4$
because of entropic repulsion between the chains. 

In this work, our aim is to extend the work presented in
Ref. \cite{Ala96} to carry out a systematic study of
the spreading and diffusion of interacting chainlike molecules on
surfaces. To begin with, we present an analysis of
density profiles of spreading droplets with qualitatively different
forms of $D_c(\theta)$. Using Fick's law, we show how the spherical
cap type of shapes result from the increase of $D_c(\theta)$ {\it vs.}
$\theta$, and then quantitatively determine $D_c(\theta)$ from the
experimental profiles of Ref. \cite{Alb92} up to $\theta \approx 0.4$.
By extensive Monte Carlo
simulations of a discrete model of interacting chainlike molecules,
we then compute $D_c(\theta)$ for all coverages, and
demonstrate that the chainlike nature of the molecules 
induces a strong repulsive interaction. When this dominates over
interchain attractions, $D_C(\theta)$ first increases as a function
of $\theta$ as deduced from the experiments, 
but eventually must go to zero for $\theta \rightarrow 1$.
The relative maximum of $D_C(\theta)$ decreases for increasing
attractions. We also examine the profiles of spreading droplets
in detail for various interactions. For the case of pure entropic
repulsion, we find excellent agreement between experimental and numerically
obtained profiles. This demonstrates that steep density profiles can
be obtained from energetic considerations
without having to assume that the film edge acts as
a phase boundary between a 2D condensate and a vapor phase, as 
suggested previously \cite{Fra93}. For increasing attractive interactions,
the corresponding droplet shapes are in between spherical cap and
Gaussian shapes.

Finally, we develop an analytic mean field 
approach to explain the behavior
of the collective diffusion coefficient for chainlike molecules 
by using the fact that $D_C(\theta)$
can be expressed as the
product of thermodynamic factor $\chi_0^{-1}$ and the mobility $m$. 
The thermodynamic factor is
evaluated using effective free energy arguments, and
the chain mobility is calculated numerically using the recently
developed dynamic mean field theory \cite{Hje97}. Good agreement
is found between theory and the numerical simulations.

The outline of the paper is as follows. In Sec. II we describe
the analysis of spreading profiles from
the diffusion equation which gives partial information about
the collective diffusion coefficient. The lattice
model used in this work is explained in Sec. III.
The results of numerical simulations are
presented in Sec. IV.  We discuss briefly the results for athermal
chains presented in Ref. \cite{Ala96}, and present complete
results for chains with attractive interchain interactions. 
In Sec. V we present the mean field theory for the
thermodynamic factor $\chi_0^{-1}$ and the mobility $m$.
In Sec. VI we use the chain-chain pair distribution function to extract
the effective pair interaction potential for athermal chains.
Finally, in Sec. VII we briefly summarize our main results.

\section{Analysis of Spreading Profiles from the Diffusion 
Equation}

As discussed in the Introduction, experimental measurements 
of spreading density profiles of a polymer film 
in the submonolayer regime can be used to
obtain information about the coverage dependence of 
the collective diffusion coefficient $D_C(\theta)$. 
This was first realized by Herminghaus
{\it et al.} \cite{Her95} who
considered the spreading of polydimethylsiloxane
(PDMS) on metal surfaces, including the data of Ref. \cite{Alb92}. 
A typical late-time spreading profile taken from this experiment
is shown in Fig. 1. The height of the profile is measured in {\AA}ngstroms 
({\AA}), but it actually describes the average areal density distribution
of polymers on the surface.
As noted in Ref. \cite{Alb92}, this and
the other submonolayer profiles can be fitted very well with spherical
cap function $\theta(x) = \theta_0+ \sqrt{r^2 - x^2}$ 
which is shown in Fig. 1 with a dotted
line ($\theta_0$ and $r$ are fitting parameters). 
In the same figure we also
show a Gaussian profile with a dashed line
for comparison. While the shape of the profile
clearly follows the spherical cap shape for most of the
range of densities, at the lowest coverages the film edge displays
a Gaussian shaped tail.

A qualitative way of analyzing the implications of the
density profiles for $D_C(\theta)$ is to apply
the non-linear diffusion equation \cite{Gom90,Her95}

\begin{equation}
\label{Eq:Fick}
\frac{\partial \theta(x,t)}{\partial t} = \frac{\partial}{\partial x}
\left[D_C(\theta) \frac{\partial \theta(x,t)}{\partial x}\right].
\end{equation}

Starting from a delta function type of initial profile, the
solution $\theta(x,t)$ of Eq. (1) is an exact Gaussian
profile for all times when $D_C = const.$ 
In Fig. 2 we show typical results from
the numerical solution of Eq. (1) for three choices of $D_C(\theta)$,
including a monotonously {\it decreasing} 
and {\it increasing} functions of coverage
$D_C(\theta)=1-0.99\theta$ and a $D_C(\theta)=0.01+0.99\theta$,
respectively. For the first case, profiles are obtained that
are {\it below} the Gaussians for high and intermediate 
coverages. On the other hand, for cases where  $D_C(\theta)$
increases, profiles are {\it above} the Gaussian solution.
This is exactly what was observed in the experiments (see Fig. 1)
and thus we can conclude that $D_C(\theta)$ must be an 
{\it increasing} function of coverage for the range
of covereges corresponding to the density profiles
(see also Ref. \cite{Her95}).

It is possible to determine $D_C(\theta)$ more quantitatively
by solving density profiles from Eq. (1) and matching them
with the experimental ones. We did this by using a 
fitting function for $D_C(\theta)$ of the form 
$D_C(\theta)=c_1+c_2\tanh[c_3(\theta-c_4)]$, where
$c_1$, $c_2$, $c_3$, and $c_4$ are fitting parameters. By adjusting
these parameters it is relatively easy to obtain profiles
with shapes closely matching the experimental ones of
Ref. \cite{Alb92}, with $c_1=0.159$, $c_2=0.144$, $c_3=5.174$,
and $c_4=0.345$. The resulting profile is shown in Fig. 1 with a solid 
line. The matching of the profiles also fixes the normalized
coverage scale from Fick's law corresponding
to the effective layer thickness in the experiments.
The analysis reveals that $D_C(\theta)$ at very low concentrations
is almost constant and then monotonically increases up to about
$\theta = 0.4$, in agreement with Ref. \cite{Her95}. 
The resulting $D_C(\theta)$ 
from the best fits is shown in the inset of Fig. 1. We not that
the behavior of $D_C(\theta)$ at higher values of $\theta$
cannot obtained from this procedure, because the maximum value of the
experimental profile at hand is already less than $\theta =0.4$.

\section{Model of Chainlike Molecules}

To determine $D_C(\theta)$ for all values of $\theta$ and
to study the effects arising from the chainlike nature of
the diffusing molecules on surfaces, we will in the present work
use a discrete model called the fluctuating bond (FB) model
\cite{Car88}. This minimal model of polymers is widely used
for simulations of many-chain systems \cite{Bin95}. The idea in
the FB model is that it is a coarse-grained model of real polymer
chains. Real polymers, such as simple polyethylene consist of 
repeated segments of ${\rm CH}_2$ monomers
where carbon atoms are bound to each other forming a 
long chain. The bond length and the bond angle between adjacent 
carbon atoms are almost fixed. However, the torsional angle between
adjacent bonds can have different values, and thus
the end-to-end distance of a long chain of ${\rm CH}_2$ segments
may vary significantly. On a coarse-grained level then, such a chain
can be described by a reduced number of effective segments.  

In the 2D version of FB model the chains in the model consist of connected 
segments
that occupy sites on a square lattice. Each segment prohibits all other
segments from occupying its nearest or next nearest neighbor
lattice sites. In the model (see Fig. 3),
the distance between adjacent segments $\ell$ 
can vary between $2\le \ell \le \sqrt{13}$ in lattice 
units, where the upper limit prevents bonds from crossing each other. 
With these restrictions there are 36 different bond vectors in the model.
The FB model has been shown to give static properties, such as pressure,
in full agreement with simulations of continuum models \cite{Deu90}.
Furthermore, the FB model incorporates the same type of dynamics for 
a single chain as the continuum Rouse model \cite{Car88}.

To consider the general case where there are attractive interactions
between the chains and the flexibility of individual chains
can also vary, we have used the following 
Hamiltonian:

\begin{equation}
\label{Eq:interactions}
\frac{{\cal H}_{\rm eff}(r,\phi)}{k_BT}=
\frac{4\epsilon}{k_BT}\sum_{l\neq l'}^n\sum_{i,i'}^{N_{FB}} 
\left[\left(\frac{\sigma}{r_{l,l';i,i'}}\right)^{12}-\left(\frac{\sigma}
{r_{l,l';i,i'}}\right)^6
\right] -\frac{E_s}{k_BT}\sum_{l=1}^n\sum_{i=1}^{N_{FB}-1} \cos(\phi_{l,i}).
\end{equation}

In Eq. (2), the first term is a Lennard-Jones type
of potential, where $J \equiv 4\epsilon/k_BT$ is the strength of the
interaction, $r_{l,l';i,i'}$ 
is the distance between segments $i$ and $i'$ 
of different chains $l$ and $l'$, and $\sigma$
was chosen such that the potential minimum was at the 
distance of two in lattice units. The cut-off radius 
of the potential was $\sqrt{10}$ lattice units, after which the
potential is small enough to give a negligible contribution.
In the summations, $n$ is the 
number of chains and $N_{FB}$ is the number of segments in each chain.
The second term controls the stiffness of each chain, with
$\phi_{l,i}$ being the angle between 
two adjacent bonds $i$ and $i+1$ in chain $l$ and
$K \equiv E_s/k_BT$ the stiffness parameter. For
increasing $E_s$ or decreasing temperature $T$, the
chains become more stiff.

Dynamics is introduced in the model by Metropolis
moves of single segments, with a probability of acceptance
$\min[e^{-\Delta {\cal H}_{\rm eff}/k_BT},1]$, 
where $\Delta {\cal H}_{\rm eff}$ is the energy difference between
final and initial configurations for acceptable moves to a nearest
neighbor site, for which
site exclusion and bond length restrictions must be satisfied.
One Monte Carlo (MC)
time step is defined as an attempt to move each segment of every chain.
In Fig. 3 we show a typical configuration of a chain 
in the FB model for $N_{FB}=6$. One possible move of a segment
is shown by the dashed line. It should be noted that 
since the dynamics within the FB model consists of single
segment moves only and there are no direct translational
modes, the limit of a rigid rod is not well defined \cite{dynamics}.

\section{Numerical Simulations}

\subsection{Athermal Chains}

To quantitatively determine the coverage dependence of
diffusion, we have performed extensive MC simulations using
the FB model explained in Sec. III. 
First, we consider the case of fully flexible, athermal chains
for which $J=K=0$ \cite{Ala96}. The linear size of the 
2D square lattice we used was $L=180$ for most cases.
To calculate $D_C(\theta)$ we used the temporal decay of the Fourier 
transformed density autocorrelation function $S(k,t)=S(k,0) e^{-k^2D_C
(\theta)t}$, where $S(k,0)$ becomes constant in the hydrodynamic limit 
$k\rightarrow 0$. The density-fluctuation autocorrelation function is 
defined as
$S(r,t)=\langle \delta \theta(r,t) \delta \theta(0,0)
\rangle$, where $\delta \theta(r,t) \equiv \theta(r,t) 
- \langle \theta(r,t) \rangle$, and $S(k,t)$ is its
Fourier transform \cite{Mak88}. 
It is calculated separately for each fixed value of 
the average normalized coverage $\theta\equiv \langle \theta(r,t)
\rangle$ which for the FB model is defined
to be $\theta=4nN_{FB}/L^2$.
In Fig. 4 we show results of these calculations for the case $N_{FB}=6$
with circles. 
The results are normalized with $D_1$, which is the diffusion constant 
of a single monomer ($N_{FB}=1$) in the limit $\theta\rightarrow 0$.
Initially, $D_C(\theta)$ increases
up to $\theta \approx 0.7$, after which it rapidly
approaches zero.
The initial behavior of $D_C(\theta)$ is in good agreement
with results obtained from the shape of the spreading profiles.

It is interesting to compare the behavior of $D_C(\theta)$
with the {\it tracer diffusion coefficient}
of individual chains $D_T(\theta)$ defined by

\begin{equation}
\label{Eq:trace}
D_T =
\lim_{t\rightarrow\infty}{1\over 4tn}\sum_{i=1}^n
\langle |\vec r_i(t)-\vec r_i(0)|^2\rangle,
\end{equation}
 
where $\vec r_i(t)$ is the position vector
of the $i^{\rm th}$ chain at time $t$.
In Fig. 5 we show the results from simulations for
the athermal case with circles. 
In the limit $\theta\rightarrow 0$ $D_C(0) = D_T(0)$,
and thus we use the same normalization as in Fig. 4.
As expected from increased interchain blocking,
$D_T(\theta)$ is a monotonically decreasing function of $\theta$.
This is a striking example of how fundamentally different the
behavior of the two types of diffusion coefficients can be.

We have also studied the effect of chain length and stiffness to
collective diffusion, and a summary of the results can be found
in Ref. \cite{Ala96}. When the length of the chains increases diffusion
slows down, but the relative maximum of $D_C(\theta)$ becomes more 
pronounced. With increasing stiffness
diffusion slows down, too, but now the maximum of $D_C(\theta)$ 
becomes less pronounced.

\subsection{Chains with Attractive Interactions}

To study the effect of attractive interactions between polymers 
we used the FB model with the Hamiltonian of Eq. (2) with two
interaction parameters, namely $J=-0.5$ and $J=-1.0$.
For the results presented here, we considered fully flexible
chains ($K=0$) of length $N_{FB}=6$.
The sizes of the lattices in the MC simulations
varied from $100 \times 100$ to $180 \times 180$. 

The results for the two cases as obtained from the
decay of $S(k,t)$ are shown in Fig. 4.
As the strength of the attractive interaction increases,
diffusion slows down but the overall behavior 
as a function of $\theta$
is qualitatively similar to the athermal case. 
The main influence of the attraction is to significantly
reduce the relative height of the maximum of $D_C(\theta)$
\cite{glassy}.

To compare with the athermal case,
we also calculated $D_T(\theta)$ for which the results
are shown in Fig. 5. When the strength of the interaction
increases, $D_T(\theta)$ decreases more rapidly as a function of 
concentration.

\subsection{Spreading of Droplets}

We also used the FB model to directly simulate 
the spreading of 2D submonolayer droplets. The chains were initially
confined to a circular region, and after equilibration
the spatial constraints were removed.
The consequent spreading was monitored and corresponding
density profiles calculated as a function of time.
We find that within the accuracy of the fits, there is a
linear relationship between the experimental and MC time scales
which supports the use of the single-segment dynamics within the
FB model.

In Fig. 6 we show a comparison between three
experimental submonolayer profiles \cite{Alb92}, profiles
obtained from the MC simulations for athermal chains,
and profiles computed using Eq. (1) with the tanh fitting
function for $D_C(\theta)$ as explained previously. 
The agreement between all the profiles is excellent
demonstrating the consistency of our approach. It also shows that
the behavior of the PDMS polymers used in the experiment 
can be most simply explained in terms of athermal chain dynamics,
with the entropic repulsion dominating.
This is in contrast to the experiments of Ref. \cite{Fra93},
where the steep film edges were assumed to be a 1D phase
boundary between a 2D condensate and a vapor phase. Our results
here demonstrate that no such assumptions need to be made; the
steep spherical cap shapes of submonolayer droplets are
expected to be 
a {\it generic} signature of strongly repulsive 
effective interactions.
 
In Figs. 7(a) and 7(b) we show additional
spreading profiles for the cases $J=-0.5$ and
$J=-1.0$, respectively. The changes seen in $D_C(\theta)$
in Fig. 4 can also be seen in the corresponding spreading profiles. 
In Figs. 7(a) and 7(b) the simulation results are
shown with circles, a spherical cap fit with a solid line and
a Gaussian fit with a dotted line. 
The main result is that with increasing attractive interactions,
the shape of profiles changes from the spherical cap
towards a Gaussian shape (except for the highest coverages).
These results show the intimate connection between $D_C(\theta)$
and spreading profiles; in the regime where $D_C(\theta)$
is almost constant, the corresponding profile shape is close
to the Gaussian limit.

\section{Mean Field Theory for Collective Diffusion}

To better understand the somewhat unusual behavior
of collective diffusion of chainlike molecules, we start
from the Green-Kubo relation \cite{Gom90}

\begin{equation}
D_C = \lim_{t\rightarrow\infty}\frac{1}{2 \langle (\delta n)^2 \rangle}
\int_0^{\infty} dt \langle \vec J(0) \cdot \vec J(t) \rangle \ ,
\end{equation}

where $n$ is the number of chains, 
$\vec J(t) = \sum_{i=1}^n 
\vec v_i(t)$ is the total particle 
current, and $\langle (\delta n)^2 \rangle$
the mean square fluctuation of chains (in a finite area $A$).
In terms of the mean square displacements of the 
individual chains, this can be written as

\begin{eqnarray}
D_C  
&=&\lim_{t\rightarrow\infty}\frac{\langle n \rangle}{\langle(\delta n)^2
\rangle}\frac{1}{4tn}\left\langle
\left(\sum_{i=1}^n\vec r_i(t)-\vec r_i(0)
\right)^2\right\rangle \nonumber\\
& \equiv & \chi_0^{-1}  m, 
\end{eqnarray}

where $\vec r_i(t)$ is the position of the center of mass of the 
$i^{\rm th}$ chain at time $t$. In this equation, the term
$\chi_0^{-1}=\langle n\rangle /\langle(\delta n)^2\rangle$ defines
the thermodynamic factor, and the remainder of the equation
defines the mobility $m$.
The thermodynamic factor is related to the density 
fluctuations of the system, while the mobility can be
written as 

\begin{equation}
m=\lim_{t\rightarrow\infty}\frac{n}{4t}
\langle(\Delta \vec r_{CM})^2\rangle,
\end{equation}

where $\Delta \vec r_{CM}=\vec r_{CM}(t)-\vec r_{CM}(0)$
is the displacement of the {\it center of mass} of all the chains, and
$\vec r_{CM}(t)\equiv (1/n) \sum_{i=1}^n
\vec r_i(t)$.
In the theory presented here, we will treat the two
factors $\chi_0^{-1}$ and $m$ separately.

\subsection{Thermodynamic Factor}

To estimate the thermodynamic factor we consider a generalization
of the simple 
thermodynamic theory presented for athermal chains in Ref.
\cite{Ala96}. We take as a starting point an effective
Helmholz free energy $F$ as

\begin{equation}
F=F_0-na\theta^b-k_BT\ln [\frac{n!}{(M-n)!n!}] -nk_BT\ln w.
\end{equation}

The first term $F_0$ is a constant, while the
second term $E_{int}=na\theta^b$ comes from attractive 
interactions between segments of different chains \cite{Gom90}, and
is temperature dependent. 
We calculated $E_{int}$
directly from the MC simulations at different coverages to estimate $a$ and 
$b$ and verified that this approximation is well satisfied.
For the cases $J=-0.5$ and $J=-1.0$ the resulting
values of these parameters 
are $a/k_BT=1.8$, $b=1.6$, and $a/k_BT=2.7$, $b=1.4$,
respectively, while for the athermal case $a=b=0$.
The third term comes from the entropy of the center of the mass of the
$n$ polymers and it is here approximated by the expression for a 2D Langmuir
gas on a lattice \cite{Gom90}. The parameter $M=L^2$ denotes 
the total number of lattice sites and thus $\theta=nN/M$
(for the FB model $\theta=4nN_{FB}/M$ due to the exclusion
rules). The last term in 
Eq. (7) is the entropic contribution from the chainlike
nature of the molecules, where $w$ is the number of
possible configurations of each chain, and is a model dependent
quantity.
For the present case of chainlike molecules, we approximate it
by decoupling the total number of configurations into the product
of two terms 

\begin{equation}
\label{Eq:w1w2}
w(\theta)=w_1^2(\theta)w_2^{N-2}(\theta),
\end{equation}

where $w_1(\theta)$ is the entropy
arising from a segment at the end of the chain,
and $w_2(\theta)$ from each segment in the middle of the chain.
For the FB model with different interactions, we have 
numerically determined $w_1(\theta)$ and $w_2(\theta)$. 
In Fig. 8(a) we show the behavior of $w_1$ and
$w_2$ with three different values of $J$,
namely $J=0$, $J=-0.5$ and $J=-1.0$. 
These quantities can be easily interpolated for all coverages
\cite{fits}.

The chemical potential $\mu$ can be calculated from $F$ by using
$\mu = (\partial F/\partial n)_{T,V}$ which gives

\begin{equation}
\frac{\mu(\theta)}{k_B T}=\frac{\mu_0}{k_B T} -a'(b+1)\theta^b + 
\left[\ln\left(\frac{\rho}{1-\rho}\right) - \ln(w)\right], 
\end{equation}

where $\rho=\theta/N$ is the number of chain molecules per unit area 
with $N$ segments, and $a'=a/k_BT$.
It can be shown that the thermodynamic factor can be written
as $\chi_0^{-1}=\theta [\partial (\mu /k_BT)/\partial\theta]$ 
\cite{Gom90}, and thus we obtain

\begin{equation}
\label{Eq:appr}
\chi_0^{-1} = 1 - a'b(b+1)\theta^b + \frac{\theta}{N-\theta} - \theta 
\frac{\partial \ln(w)} {\partial \theta}.
\end{equation}

In Fig. 8(b) the markers show results for $\chi_0^{-1}(\theta)$ 
as obtained from accurate MC simulations of
the density fluctuations of the FB model
(with $N_{FB}=6$) in equilibrium, as extrapolated
to an infinite system size. At very high concentrations
density fluctuations are so small that it is very difficult to
obtain accurate results. Furthermore, with attractive interactions
($J\neq 0$) for coverages
$\theta \gsim 0.8$ the dynamics of the system slows down \cite{glassy}
and thus we here present results only for 
smaller coverages. 
Using the approximation of Eq. (\ref{Eq:appr}) with numerically
determined $w_1(\theta)$ and 
$w_2(\theta)$ of Fig. 8(a) with $N_{FB}=N$
for the athermal case, the results show that
the true magnitude of $\chi_0^{-1}(\theta)$ is somewhat
underestimated throughout the range of coverages \cite{Ala96}.
However, if the magnitude of the thermodynamic factor
is known for some coverages, the effective chain length $N$
appearing in Eq. (\ref{Eq:w1w2}) can be used as an additional
fitting parameter to improve the results. In Fig. 8(b)
we show the results of this approach, with only one 
parameter $N$ fitted to our
MC data for $\chi_0^{-1}(\theta)$.
The corresponding values of the parameters for
$J=-0.5$ and $J=-1.0$ are $N=11$, and  $N=6$, respectively. 
For the athermal case $N=16$ gives the best results.
Thus, for increasing attraction, the approximation 
of Eq. (\ref{Eq:w1w2}) seems to become more accurate.

\subsection{Calculation of Mobility}

While the thermodynamic factor contains information about the
equilibrium density fluctuations, the mobility $m$ 
is determined by the dynamics of the center-of-mass motion
of the particles. 
To calculate $m$ theoretically we use the recently developed
Dynamical Mean Field (DMF) theory \cite{Hje97}, which 
yields an approximate expression for $m$ as

\begin{equation}
m_{\rm DMF} = \frac{\ell^2}{4}\Gamma, 
\label{Eq:collective}
\end{equation}

where $\ell$ is the effective jump length and $\Gamma$ is the average
jump rate. This formulation makes
it very efficient to numerically evaluate $m$, and has recently
been shown to give a very good approximation of the true $m$
for various strongly interacting systems \cite{Hje97}.

In Fig. 9(a) we show mobilities calculated from MC 
simulations of $m(\theta)$ for the FB model for the cases $J=0$, 
$J=-0.5$ and  $J=-1.0$, using the definition of Eq. (6). 
The data are normalized with the mobility of one segment $m_1$.
In the same figure we show also the results calculated from
the DMF theory of 
Eq. (\ref{Eq:collective}). The effective jump length $\ell$ has been estimated 
from the zero coverage limit, where $\chi_0(0)=1$ and
thus $m_1=D_1$.
In the athermal case the DMF theory is in good agreement with 
simulation results, but with increasing attraction between 
the chains it starts to deviate more 
from the MC simulation results. This behavior
is physically reasonable, because attractive interactions
strengthen the effect of 
dynamical correlations that are not included in the DMF theory
\cite{Hje97}. Despite this,
DMF gives the qualitative behavior of $m(\theta)$ 
rather well even in case of attractive chains.

A commonly used method to approximate the mobility is based on
the Darken equation \cite{Gom90}. It states that the mobility can be 
approximated by the tracer
diffusion coefficient, {\it i.e.} $m(\theta) \approx D_T(\theta)$.
In Fig. 9(b) we show the complete results from 
simulations of $D_T(\theta)$ with the  
interaction parameters $J=0$, $J=-0.5$ and $J=-1.0$. 
In same figure, the markers denote
the results of direct MC simulations of $m(\theta)$.
As can be seen from the comparison, for the chainlike
molecules $D_T(\theta)$ and $m(\theta)$ behave quite
differently; for athermal chains even the curvatures
of the two functions have opposite signs. 

\section{Effective Potential}

The fact that the initial increase of $D_C(\theta)$ is due to
effective repulsive interactions between the molecules
can be seen in Eq. (10), where the important term 
$1 - \theta[\partial \ln(w)/\partial \theta]$ comes from 
entropic origin of the chainlike molecules
\cite{Her95,Ala96}. With increasing attraction,
this entropy-induced repulsion is compensated by the attractive
$E_{int}$, and the maximum value of $D_C(\theta)$ is reduced in magnitude. 

To study the effective potential corresponding to
the entropic repulsion we calculated
the chain-chain pair distribution function 
$g(r)=\sum_{ii'}^{'}\langle\delta (r-r_i)\delta (r_{i'})\rangle $ 
\cite{Boo91}, where $r_i$ is the position of the center of mass 
of chain $i$ and prime indicates that terms with $i=i'$ 
are to be omitted. From $g(r)$ we extracted numerically 
the effective pair interaction potential $V_e(r)/k_BT$ for athermal 
chains. This can be done by 
first calculating the direct correlation function $c(r)$ from the 
Ornstein-Zernike relation \cite{Ver68}

\begin{equation}
\label{Eq:Ornstein}
h(r)=c(r)+\theta\int h(r')c(|r-r'|)dr',
\end{equation}

where $h(r)=g(r) -1$ and $r$ is distance in 
lattice units. When $c(r)$ is known, 
$V_e(r)/k_BT$ can be calculated by using the hypernetted chain theory
\cite{Han76}

\begin{equation}
\label{Eq:HNC}
c(r)=-V_e(r)/k_BT+h(r)-\log[h(r)+1].
\end{equation}

In Fig. 10  we show $V_e(r)/k_BT$ for the athermal case and
for $\theta=0.25$, which shows a strong repulsion 
extending up to several
lattice cites \cite{repulsion}. 
As a comparison in the same figure there is also 
a typical Lennard-Jones potential which is more repulsive at
small distances. An interesting result of the analysis is that
for athermal chains, all the computed pair correlation
functions for coverages $\theta\leq 0.7$ and for several chain lengths
\cite{Ala96} collapse to a single function which is given by $g(\tilde {r})
=G(r\theta^{\alpha}/N_{FB}^{\beta})$, where $\alpha \approx 0.38$ and
$\beta \approx 0.55$. These scaled correlation functions 
are shown in the inset of Fig. 10.

\section{Discussion and Conclusions}

In this paper we have presented a systematic study of 
diffusion and spreading of chainlike molecules, in part
inspired by the non-Gaussian submonolayer film profiles 
observed in most experiments. Using Monte Carlo simulations
with the fluctuating bond model, we have calculated
the diffusion coefficients as a function of coverage,
generalizing the results for athermal chains of Ref. \cite{Ala96}
to chains with attractive interchain interactions.
Typically, the collective
diffusion coefficient $D_C(\theta)$ increases initially
and displays a maximum around $\theta \approx 0.7$.
The strength of the peak decreases with increasing attraction.

We have also developed a 
mean field approximation 
for the thermodynamic factor in $D_C$, while the mobility
is estimated numerically from the Dynamical Mean Field theory.
The theory reveals that the behavior observed in $D_C(\theta)$ 
is due to an entropy-induced repulsive interaction. We also extract
this interaction numerically from the pair correlation
functions for athermal chains. It is interesting to note
that the diffusive dynamics of polymers in 2D is fundamentally
different from the 3D case, where entanglement effects dominate
in dense melts of longer chains \cite{Doi92}.

The functional dependence of $D_C(\theta)$ on the coverage has interesting
consequences for the profiles of spreading films in the submonolayer
regime. When the entropy-generated repulsive interactions dominate,
the droplets assume a spherical cap type of steep shape.
Contrary to the suggestion of Ref. \cite{Fra93},
no assumptions about the film edge being a phase boundary between
a condensate and vapor need to be evoked here.
With increasing
attractive interactions these shapes evolve towars the Gaussian shape. If
these interactions dominate and $D_C(\theta)$ is a deacreasing
function of $\theta$, profiles emerge that are narrower
than Gaussians. Thus, the submonolayer spreading experiments
constitute a sensitive measure on the role of interactions in
the diffusive dynamics of polymers on surfaces.

Acknowledgments: This work has in part been supported by the Academy
                  of Finland.

\newpage

\begin{figure}[htb]
        \mbox{\hspace{4.0cm}}\begin{minipage}{12.0cm}
        \epsfxsize=12.0cm
        \hspace{-6.0cm}\epsfbox[0 300 512 620]{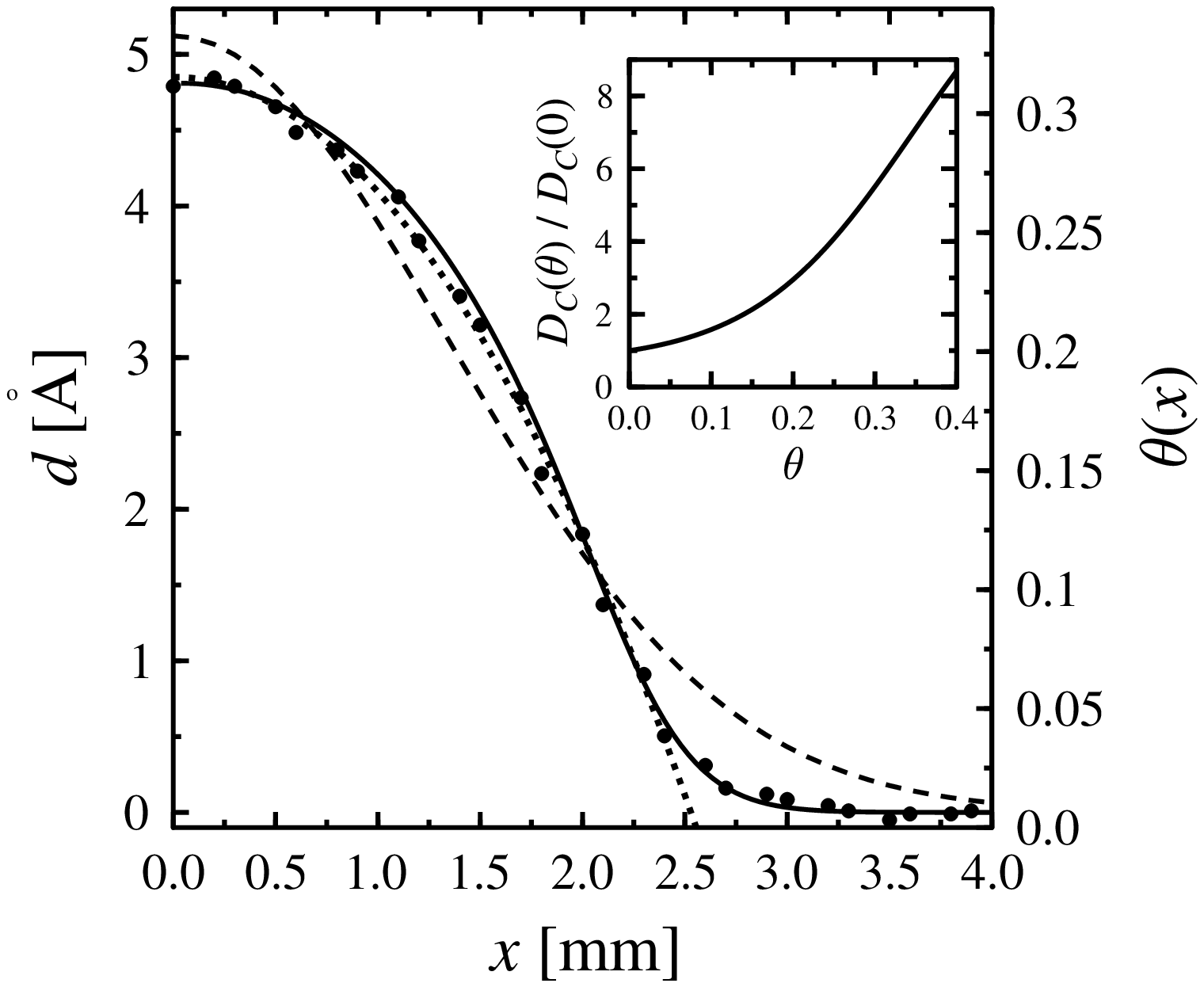}
        \end{minipage}
        \vspace{1.5cm}
        \caption{A typical late-time submonolayer spreading profile
from experiments of PDMS spreading on a silver surface \protect\cite{Alb92}
(filled circles). The height of the experimental profile is measured
in {\AA}ngstroms ({\AA}).
Dashed and dotted lines are Gaussian and spherical cap
fits, respectively. Solid line shows the solution of
Eq. (1) using $D_C(\theta)$ shown in the inset. $D_C(\theta)$ has
been normalized by the diffusion coefficient in the zero coverage
limit $D_C(0)$. 
	}
        \label{Figure2}
\end{figure}

\begin{figure}[htb]
        \mbox{\hspace{4.0cm}}\begin{minipage}{12.0cm}
        \epsfxsize=12.0cm
        \hspace{-6.0cm}\epsfbox[0 300 512 620]{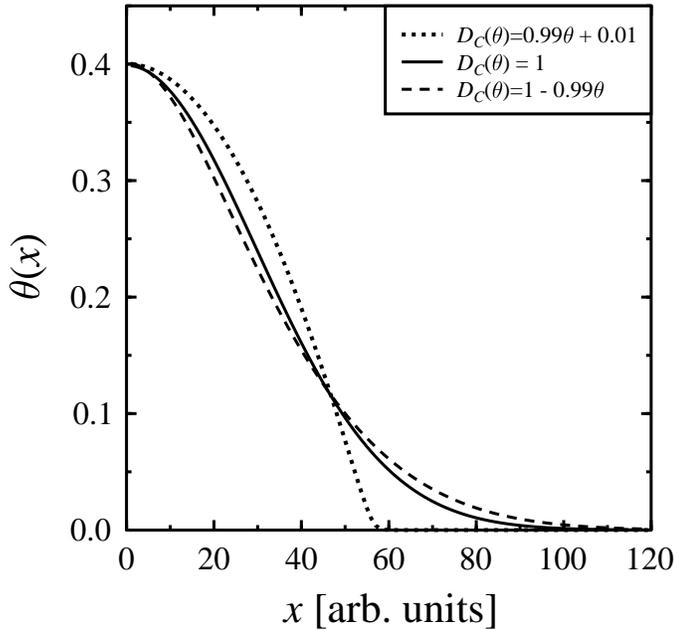}
        \end{minipage}
        \vspace{1.5cm}
        \caption{Typical results for the density profile of a
droplet from numerical solutions of Eq. (1)
with different choices of the function $D_C(\theta)$.
	}
        \label{Figure3}
\end{figure}
\newpage

\begin{figure}[htb]
        \mbox{\hspace{4.0cm}}\begin{minipage}{12.0cm}
        \epsfxsize=12.0cm
        \hspace{-6.0cm}\epsfbox[0 300 512 620]{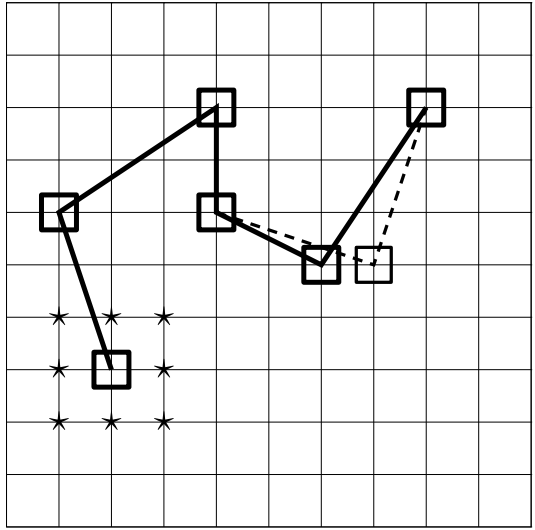}
        \end{minipage}
        \vspace{1.5cm}
        \caption{A typical configuration of a chain in the FB model with
$N_{FB}=6$. The segments are shown by squares. Stars denote the sites 
blocked by the lowermost segment.
An allowed move of one of the segments is shown by a dashed line.
	}
        \label{Figure1}
\end{figure}

\begin{figure}[htb]
        \mbox{\hspace{4.0cm}}\begin{minipage}{12.0cm}
        \epsfxsize=12.0cm
        \hspace{-6.0cm}\epsfbox[0 300 512 620]{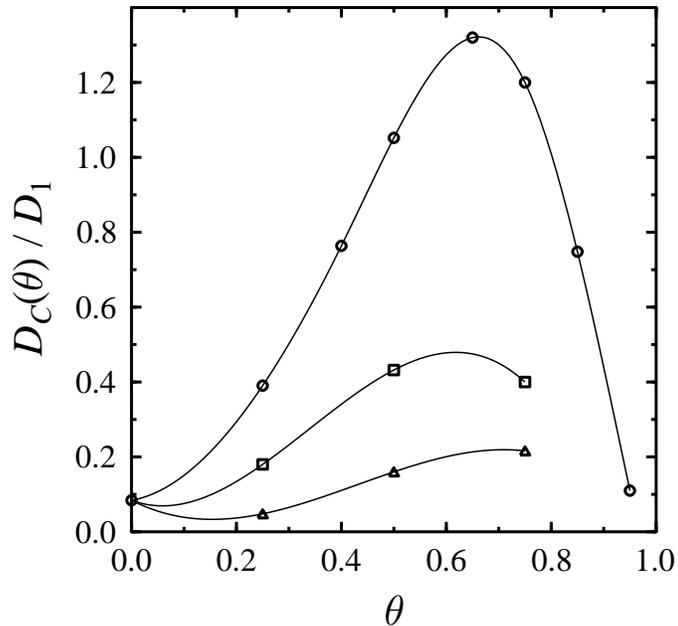}
        \end{minipage}
        \vspace{1.5cm}
        \caption{$D_C(\theta)$ for three different values of $J$
as computed from MC simulations of the FB model. Circles
are for $J=0$, squares for $J=-0.5$, and triangles for $J=-1.0$. 
Solid lines are only guides to the eye. The curves are normalized by 
the single monomer diffusion coefficient $D_1$ in the zero coverage limit.
The error bars are of the size of the symbols, or smaller.	}
        \label{Figure4}
\end{figure}
\newpage

\begin{figure}[htb]
        \mbox{\hspace{4.0cm}}\begin{minipage}{12.0cm}
        \epsfxsize=12.0cm
        \hspace{-6.0cm}\epsfbox[0 300 512 620]{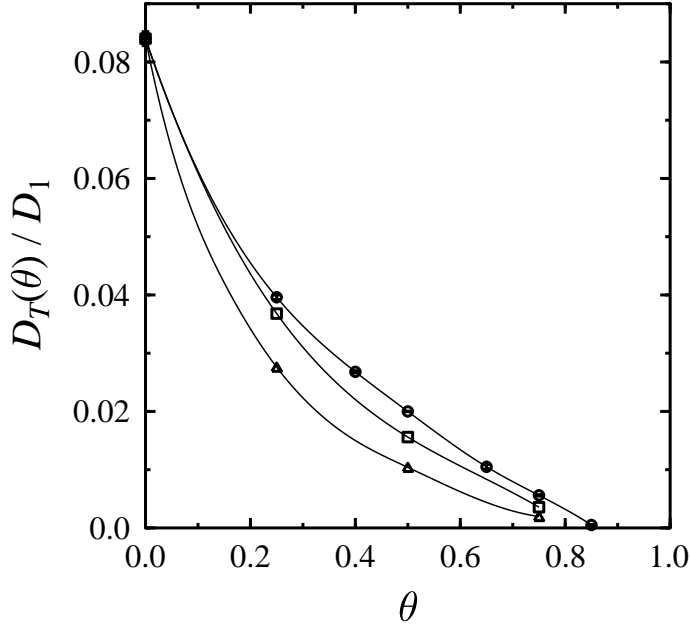}
        \end{minipage}
        \vspace{1.5cm}
        \caption{The tracer diffusion coefficient $D_T(\theta)$ for three 
different values of $J$. Circles
are for $J=0$, squares for $J=-0.5$, and triangles for $J=-1.0$. 
Solid lines are only guides to the eye. The curves are normalized by the 
single monomer diffusion coefficient $D_1$. The error bars are smaller
than the size of the symbols.	}
        \label{Figure5}
\end{figure}

\begin{figure}[htb]
        \mbox{\hspace{4.0cm}}\begin{minipage}{12.0cm}
        \epsfxsize=12.0cm
        \hspace{-5.0cm}\epsfbox[0 300 512 620]{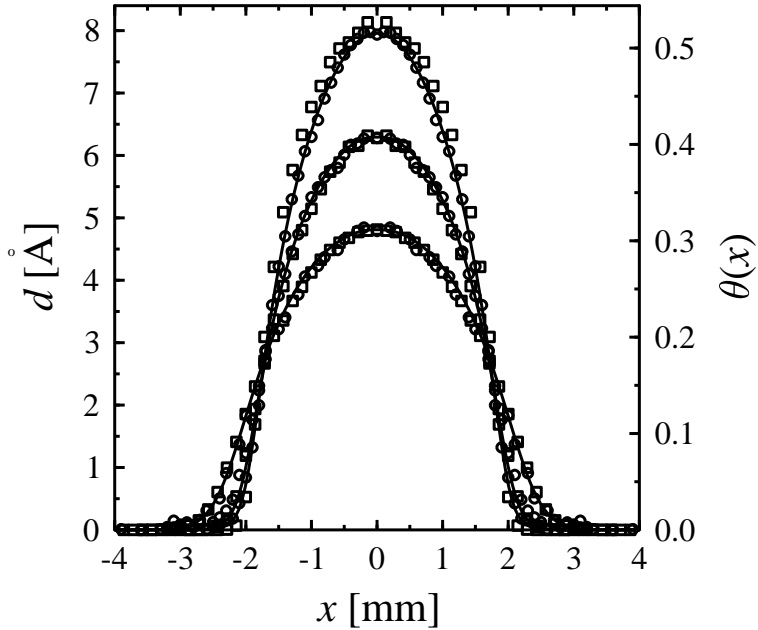}
        \end{minipage}
        \vspace{1.5cm}
        \caption{A comparison between three 
experimental submonolayer profiles measured after 50, 80, and 150 minutes
following deposition of PDMS on silver 
\protect\cite{Alb92,Ala96} (circles), MC simulations of 2D circular droplets 
from the FB model with $N_{FB}=6$ (squares), and numerical solutions of the 
nonlinear diffusion equation (solid lines). See text for details. 
	}
        \label{Figure6}
\end{figure}
\newpage

\begin{figure}[htb]
        \mbox{\hspace{4.0cm}}\begin{minipage}{11.0cm}
        \epsfxsize=11.0cm
        \hspace{-7.7cm}\epsfbox[0 300 512 620]{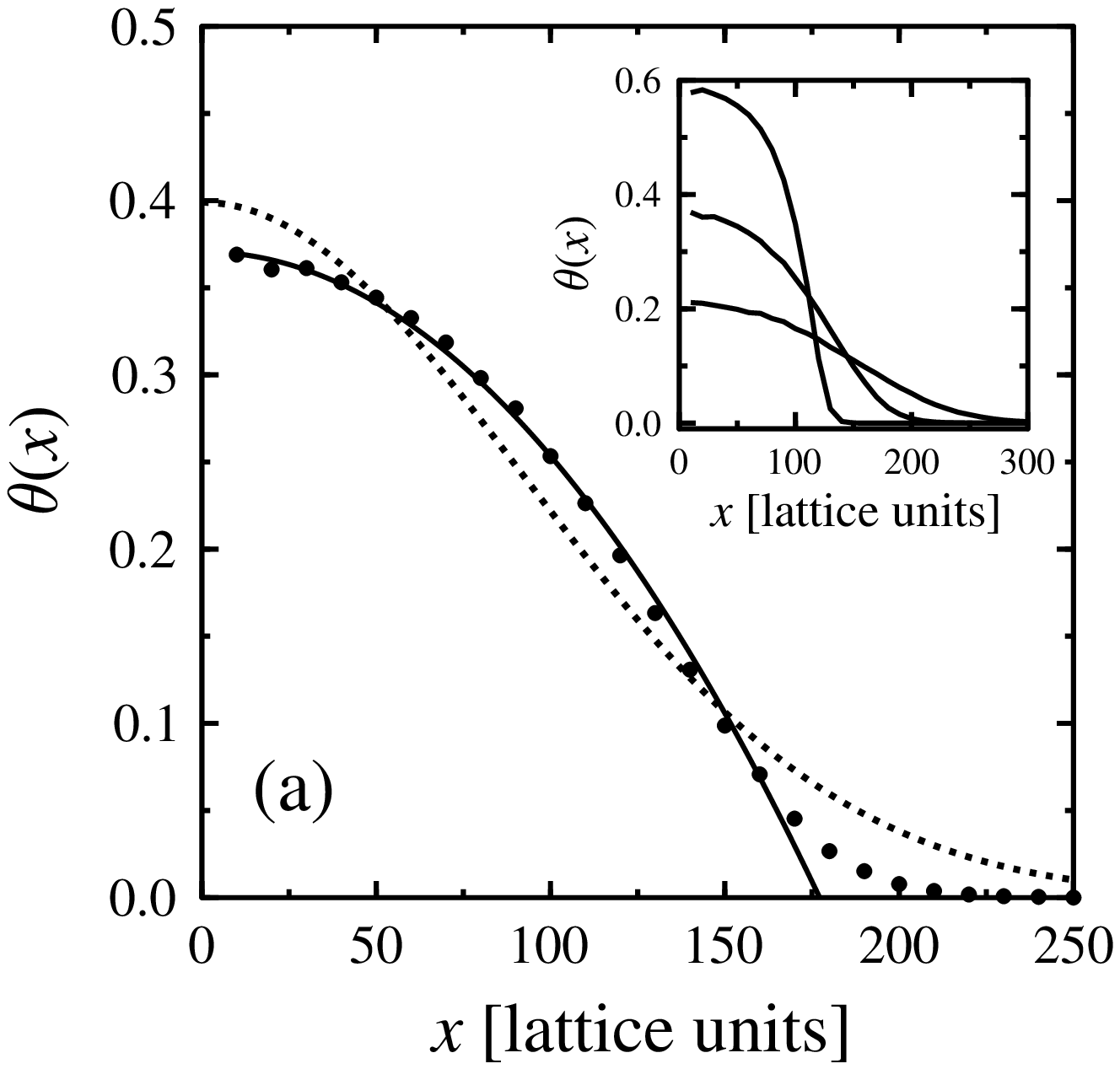}
	\end{minipage} 
	\begin{minipage}{22.0cm}
        \vspace{-5.0cm}
        \hspace{+5.5cm}
        \epsfxsize=11.0cm
	\epsfbox[0 300 512 532]{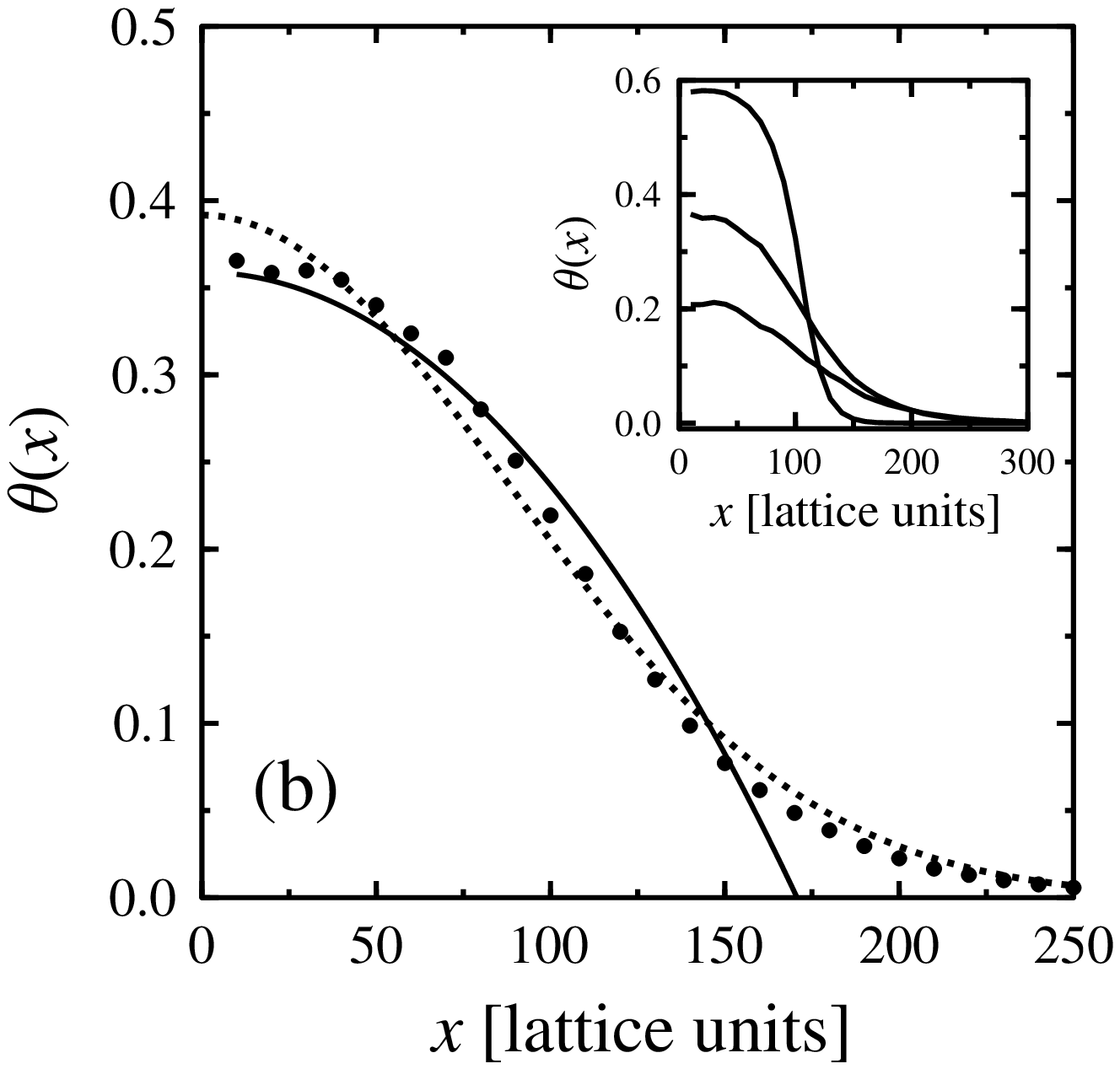}
        \end{minipage}
        \vspace{1.0cm}
        \caption{Typical spreading profiles obtained from the MC
simulations with (a) $J=-0.5$ and (b) $J=-1.0$ circles. Solid and 
dotted lines denote spherical cap 
and Gaussian fits, respectively. The insets show simulated spreading profiles 
at three different times.
	}
        \label{Figure7}
\end{figure}

\begin{figure}[htb]
        \mbox{\hspace{4.0cm}}\begin{minipage}{11.0cm}
        \epsfxsize=11.0cm
        \hspace{-7.7cm}\epsfbox[0 300 512 620]{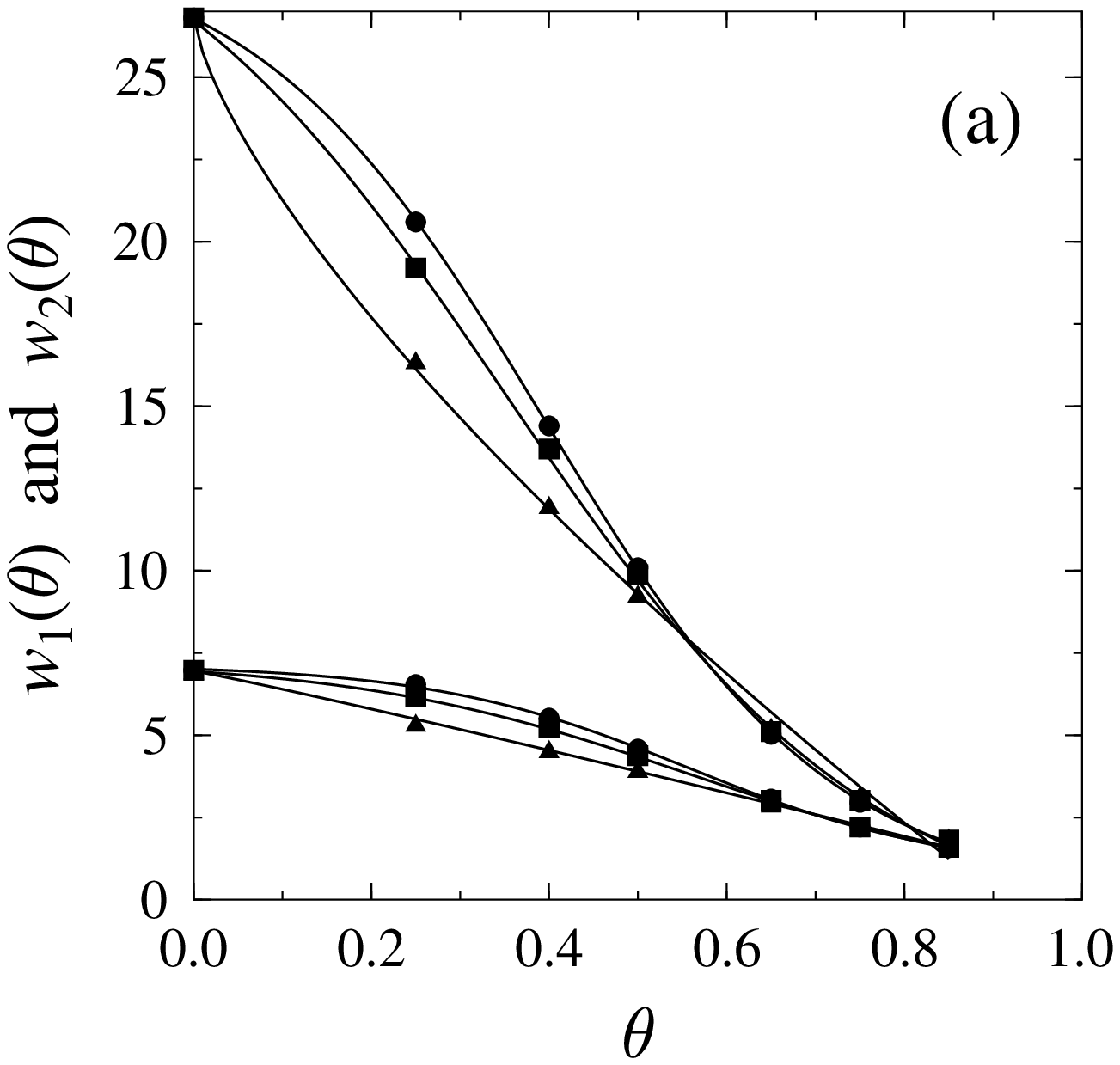}
	\end{minipage} 
	\begin{minipage}{22.0cm}
        \vspace{-5.0cm}
        \hspace{+5.5cm}
        \epsfxsize=11.0cm
	\epsfbox[0 300 512 532]{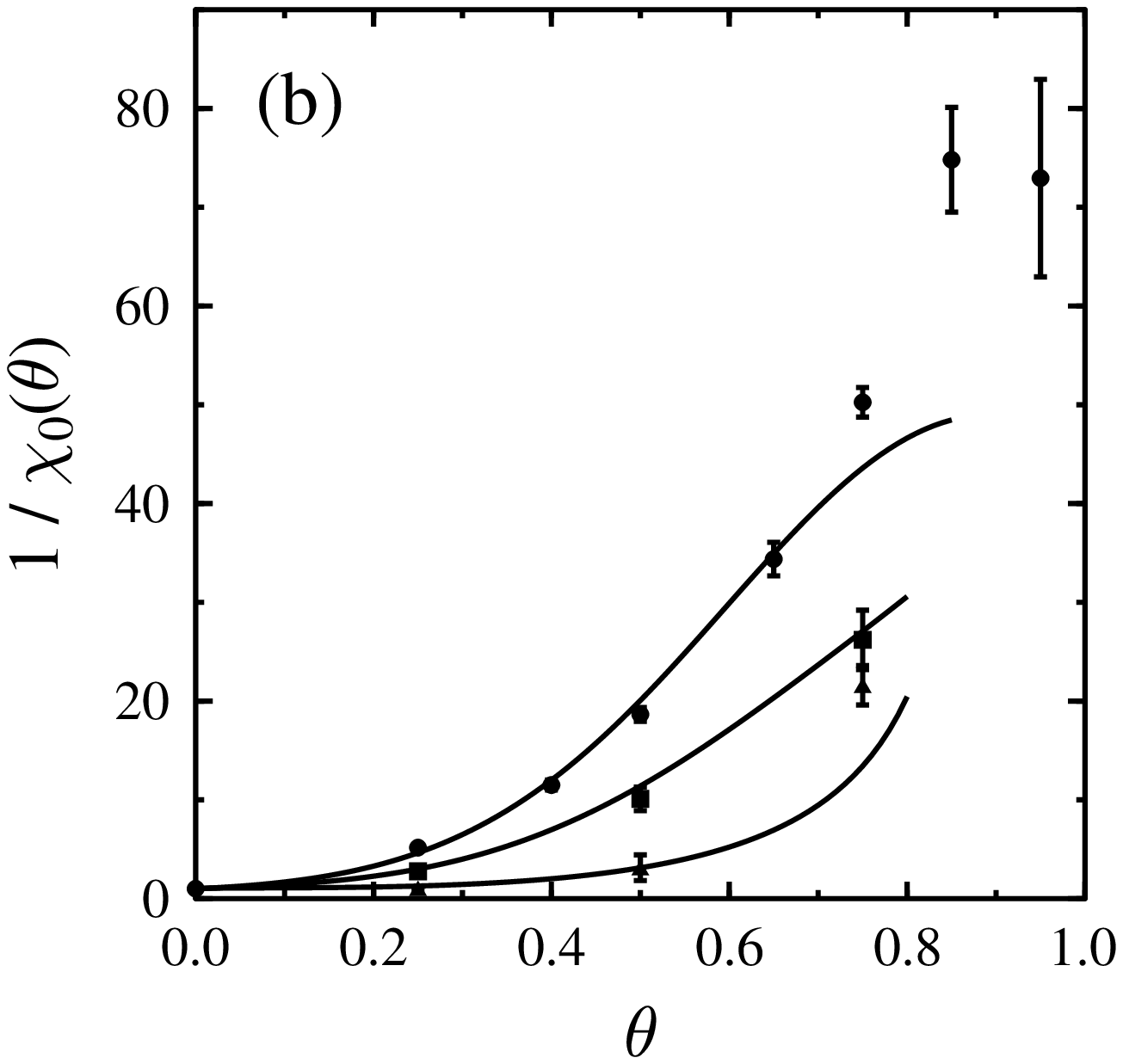}
        \end{minipage}
        \vspace{1.0cm}
        \caption{(a) Coverage dependence of $w_1(\theta)$ 
(three upper curves)
and $w_2(\theta)$ (three lower curves) for $J=0$ (circles),
$J=-0.5$ (squares), and $J=-1.0$ (triangles). Results of fitting 
\protect\cite{fits} are shown by solid lines. In the zero coverage limit,
both quantities have been computed numerically for a single
fluctuating chain and thus do not depend on the value of $J$. 
(b) Calculated coverage
dependence of $\chi_0^{-1}(\theta)$ for $J=0$ (circles),
$J=-0.5$ (squares), and $J=-1.0$ (triangles). Solid lines show
results of numerical calculations using Eqns. (8) and (10).
	}
        \label{Figure8}
\end{figure}
\newpage

\begin{figure}[htb]
        \mbox{\hspace{4.0cm}}\begin{minipage}{11.0cm}
        \epsfxsize=11.0cm
        \hspace{-7.7cm}\epsfbox[0 300 512 620]{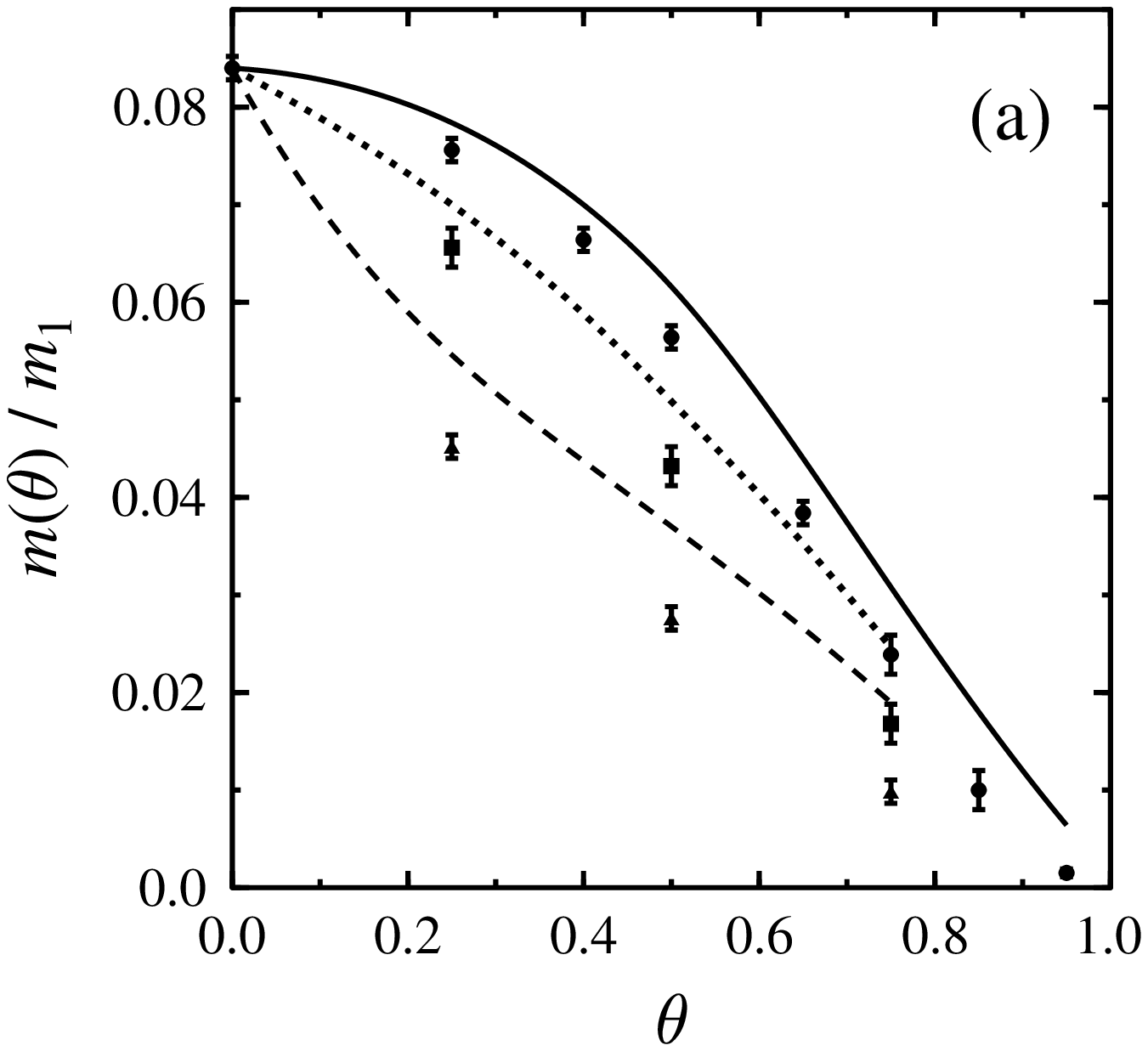}
	\end{minipage} 
	\begin{minipage}{22.0cm}
        \vspace{-5.0cm}
        \hspace{+5.5cm}
        \epsfxsize=11.0cm
	\epsfbox[0 300 512 532]{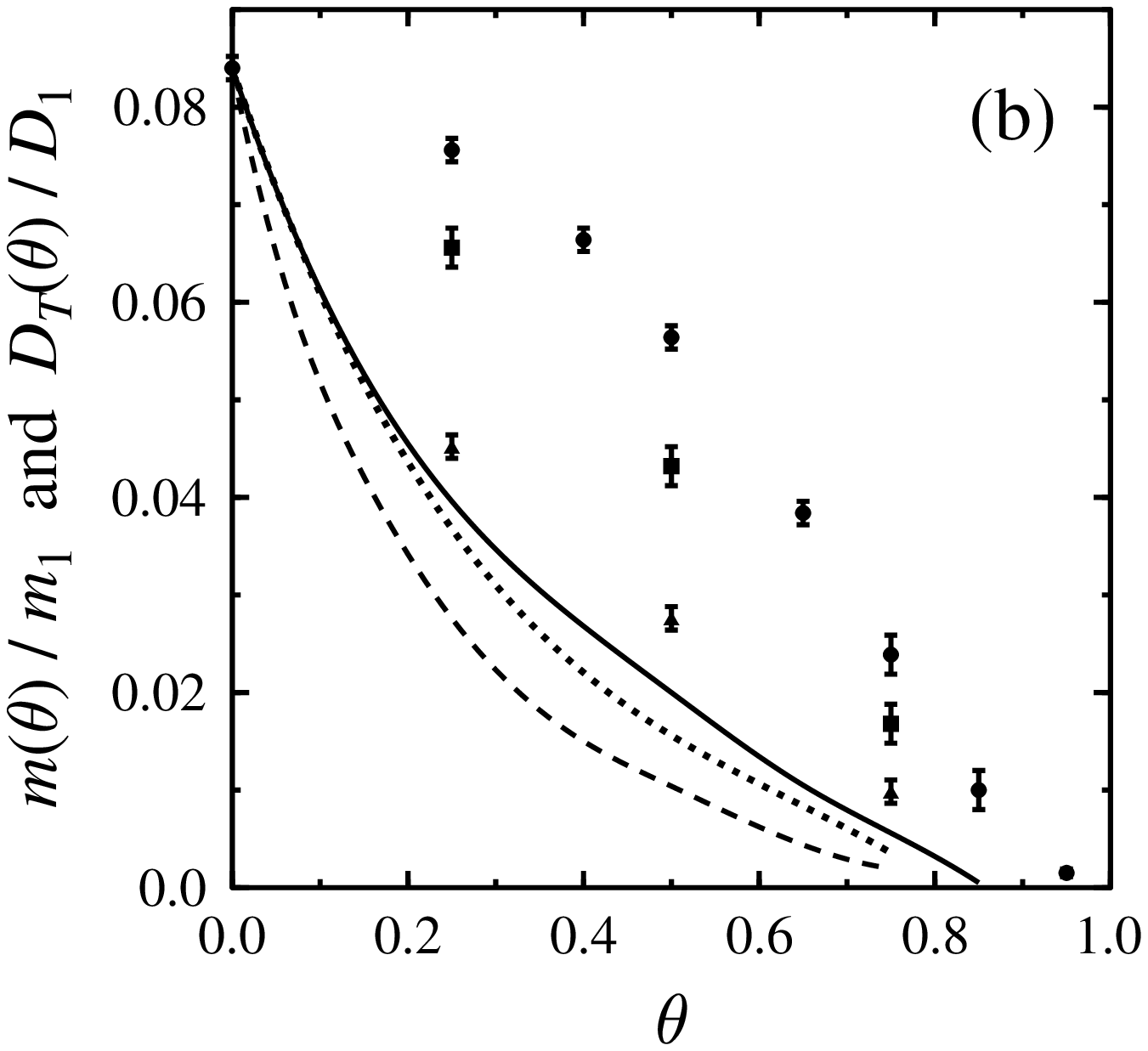}
        \end{minipage}
        \vspace{1.0cm}
        \caption{(a) Coverage dependence of $m(\theta)$ 
for $J=0$ (circles), $J=-0.5$ (squares), and $J=-1.0$ (triangles).
The corresponding results from the DMF theory are shown by solid, dotted,
and dashed lines, respectively. The data have been normalized by single
segment mobility $m_1$ in the zero coverage limit.
(b) Comparison between
$m(\theta)$ and $D_T(\theta)$ with $m(\theta)$ plotted as in (a) and 
$D_T(\theta)$ shown for $J=0$ (solid line),
$J=-0.5$ (dotted line), and $J=-1.0$ (dashed line).
	}
        \label{Figure9}
\end{figure}

\begin{figure}[htb]
        \mbox{\hspace{4.0cm}}\begin{minipage}{12.0cm}
        \epsfxsize=12.0cm
        \hspace{-6.0cm}\epsfbox[0 300 512 620]{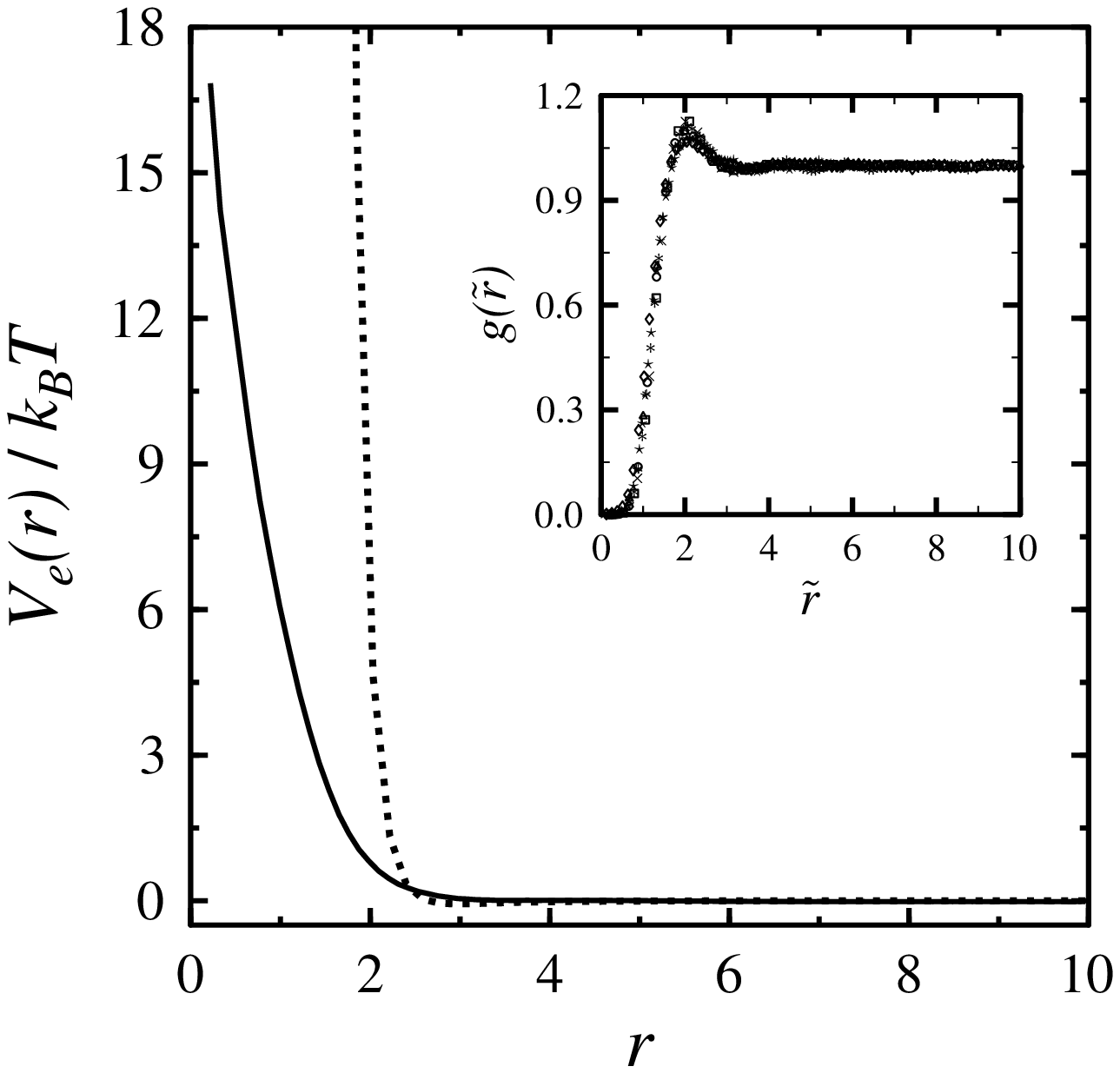}
        \end{minipage}
        \vspace{1.0cm}
        \caption{Solid line shows the effective potential $V_e(r)/k_BT$ 
between
the athermal chains in the FB model (see text for details).
The dotted line denotes
a typical Lennard-Jones potential for comparison. The inset shows 
seven scaled correlation functions $g(\tilde {r})$ for 
$N_{FB}=6$ and $\theta=0.25, \  0.40, \  0.50
\  {\rm and} \  0.75$, and for $N_{FB}=12$ and $\theta=0.25, \  
{\rm and} \  0.50$, and for $N=48$ the case $\theta = 0.25$.
See text for details. 
	}
        \label{Figure10}
\end{figure}


\begin{thebibliography}{100}

\bibitem{Gom90}
R. Gomer, Rep. Prog. Phys {\bf 53}, 917 (1990).

\bibitem{Ala92}
T. Ala-Nissila and S. C. Ying, Prog. Surface Sci. {\bf 39}, 227 (1992).

\bibitem{George}
M. V. Arena, E. D. Westre, and S. M.
George, J. Chem. Phys. {\bf 94}, 4001 (1991), and
references therein.

\bibitem{Coh92}
D. Cohen and Y. Zeiri, J. Chem. Phys. {\bf 97}, 1531
(1992); Surface Sci. {\bf 274}, 173 (1992).

\bibitem{Fic94}
K. A. Fichthorn, G. B. Prakash, and Y. Chen, Surface Sci. 
{\bf 317}, 37 (1994); D. Huang, Y. Chen, and, K. A. Fichthorn,
J. Chem. Phys. {\bf 101}, 11021 (1994);  K. A. Fichthorn,
Adsorption {2}, 77 (1996).

\bibitem{Hes89}
F. Heslot, N. Fraysse, and A. M. Cazabat, Nature {\bf 338},
640 (1989); A. M. Cazabat, N. Fraysse, F. Heslot, P. Levinson,
J. Marsh, F. Tiberg, and M. P. Valignat,
Adv. Coll. Int. Sci. {\bf 48}, 1 (1994).
 
\bibitem{Caz90}
A. M. Cazabat, N. Fraysse, F. Heslot, and P. Carles,
J. Phys. Chem. {\bf 94}, 7581 (1990).
 
\bibitem{Alb92}
U. Albrecht, A. Otto, and P. Leiderer, Phys. Rev. Lett.
{\bf 68}, 3192 (1992); Surface Sci. {\bf 283}, 383 (1993).
 
\bibitem{Fra93}
N. Fraysse, M. P. Valignat, A. M. Cazabat, F. Heslot, and
P. Levinson, J. Coll. Int. Sci. {\bf 158}, 27 (1993).
 
\bibitem{Haa95}
M. Haataja, J. A. Nieminen, and T. Ala-Nissila, Phys. Rev.
E {\bf 52}, R2165 (1995), and references therein.
 
\bibitem{Her95}
S. Herminghaus, U. Sigel, U. Albrecht, and P. Leiderer, in
{\it Proceedings of the XVth Moriond Workshop ``Short and
Long Chains at Interfaces''}, ed. by J. Daillont {\it et al.},
Condensed Matter Physics Series, Villars-sur-Ollon, Switzerland,
Jan. 21--28, 1995.

\bibitem{Ala96}
T. Ala-Nissila,  S. Herminghaus,  T. Hjelt, and  P. Leiderer, 
Phys. Rev. Lett. {\bf  76}, 4003 (1996).

\bibitem{Hje97}
T. Hjelt, I. Vattulainen, J. Merikoski, T. Ala-Nissila,  and S.C. Ying,
Surf. Sci. {\bf 380}, L501 (1997).

\bibitem{Car88}
I. Carmesin and K. Kremer, Macromolecules {\bf 21}, 2819 (1988).
 
\bibitem{Bin95}
K. Binder, {\it Monte Carlo and Molecular Dynamics Simulations
in Polymer Science} (Oxford University Press Inc., New York, 1995).

\bibitem{Deu90}
H. P. Deutsch and R. Dickman, J. Chem. Phys. {\bf 93}, 8983 (1990).

\bibitem{dynamics}
As far as single chain dynamics is concerned, recent
numerical studies using molecular dynamics 
\cite{Coh92,Fic94} reveal that diffusion of long chains cannot always
be modeled as a single segment jumps from one site to another.

\bibitem{Mak88}
C. H. Mak, H. C. Andersen, and S. M. George, J. Chem. Phys.
{\bf 88}, 4052 (1988).
 
\bibitem{glassy}
For the attractive cases, the chain configurations at coverages
close to unity freeze into a glassy state and thus it becomes
practically impossible to study diffusion with the present model.

\bibitem{fits}
For the cases $J=0$ and $J=-0.5$ we used the fitting function 
$c_1-c_2 \tanh
[c_3(c_4-\theta)]$ and for $J=-1.0$ we used 
$\tilde c_1 (1- \tilde c_2\theta^{\tilde c_3})$. 
The corresponding parameters
for the end segments for the cases $J=0$, $J=-0.5$ and $J=-1.0$ were:
$c_1=14.7, \ c_2=14.4, \ c_3=3.1, \ c_4=0.39$; 
$c_1=15.4, \ c_2=16.5, \ c_3=2.4, \ c_4=0.35$; and
$\tilde c_1=26.9, \ \tilde c_2=1.1, \ \tilde c_3=0.71$, respectively. 
The corresponding parameters for
mid-segments were: $c_1=3.9, \ c_2=3.4, \ c_3=2.9, \ c_4=0.56$; 
$c_1=3.8, \ c_2=3.5,\ c_3=2.6, \ c_4=0.56$; and $\tilde c_1=7.0, 
\ \tilde c_2=0.92, \ \tilde c_3=1.1$.

\bibitem{Boo91}
J. P. Boon and S. Yip, {\it Molecular Hydrodynamics} (Dover Publications 
Inc, New York, 1991).

\bibitem{Ver68}
L. Verlet, Phys. Rev. {\bf 165}, 201 (1968).

\bibitem{Han76}
J. P. Hansen and I. R. McDonald, {\it Theory of Simple Liquids}
(Academic Press Inc., London, 1976).
 
\bibitem{repulsion}
The fact that direct repulsive interactions between {\it pointlike}
particles typically cause $D_C(\theta)$ to increase as a function of
$\theta$ is well documented in the literature; see {\it e.g.} Refs.
\cite{Gom90,Ala92}.

\bibitem{Doi92}
M. Doi and S. F. Edwards, {\it The Theory of Polymer Dynamics}
(Clarendon Press, Oxford, 1992).

\end{thebibliography}
\end{document}